\documentclass[
superscriptaddress,
twocolumn,
amsmath,amssymb,
aps,
prl,
longbibliography
]{revtex4-2}

\usepackage{graphicx}
\usepackage{dcolumn}
\usepackage{xcolor}
\usepackage{bm}
\usepackage{physics}
\usepackage{hyperref}
\usepackage{comment}
\hypersetup{
	colorlinks = true,
	urlcolor   = blue,
	linkcolor  = blue,
	citecolor  = red
}
\usepackage{algorithm}
\usepackage{algpseudocode}
\usepackage{float}

\makeatletter
\let\newfloat\newfloat@ltx
\makeatother

\begin{document}

\title{Quasiprobability distributions with weak measurements}

\author{G. Bizzarri}
\altaffiliation{These authors contributed equally to this work}
\affiliation{Dipartimento di Scienze, Universit\`a degli Studi Roma Tre, Via della Vasca Navale, 84, 00146 Rome, Italy}
\author{S. Gherardini}
\altaffiliation{These authors contributed equally to this work}
\affiliation{Istituto Nazionale di Ottica - CNR, Largo E. Fermi 6, 50125 Florence, Italy}
\affiliation{European Laboratory for Non-linear Spectroscopy,
Universit\`a degli Studi di Firenze, 50019 Sesto Fiorentino, Italy.}
\author{M. Manrique}
\affiliation{Dipartimento di Scienze, Universit\`a degli Studi Roma Tre, Via della Vasca Navale, 84, 00146 Rome, Italy}
\author{F. Bruni}
\affiliation{Dipartimento di Scienze, Universit\`a degli Studi Roma Tre, Via della Vasca Navale, 84, 00146 Rome, Italy}
\author{I. Gianani}
\affiliation{Dipartimento di Scienze, Universit\`a degli Studi Roma Tre, Via della Vasca Navale, 84, 00146 Rome, Italy}
\author{M. Barbieri}
\affiliation{Dipartimento di Scienze, Universit\`a degli Studi Roma Tre, Via della Vasca Navale, 84, 00146 Rome, Italy}
\affiliation{Istituto Nazionale di Ottica - CNR, Largo E. Fermi 6, 50125 Florence, Italy}


\begin{abstract}
We discuss and experimentally demonstrate the role of quantum coherence in a sequence of two measurements collected at different times using weak measurements. For this purpose, we have realized a weak-sequential measurement protocol with photonic qubits, where the first measurement is carried out as a positive operator-valued measure, whereas the second one is a projective operation. We determine the quasiprobability distributions associated to this procedure using both the commensurate and the Margenau-Hill quasiprobabilities approaches. By tuning the weak measurements, we obtain a quasidistribution that may or may not exhibit negative parts, depending on the suitability of a contextual model for describing the experiment. Our results show how quasidistributions may find application in inspecting quantum monitoring, when part of the initial quantum coherence needs to be preserved.
\end{abstract}

\maketitle

\section{Introduction}

The standard description of a quantum state adopts a wave-function or a density matrix~\cite{VonNeumann}. Although their ontological status is disputed~\cite{PhysRevLett.108.150402,PhysRevLett.109.150404,Pusey12,Ringbauer15,Frauchiger2018}, they offer the most complete physical insight quantum mechanics is able to provide in its standard form. By means of Born's rule one can derive proper probability distributions associated with the measurement at single times of any physical observable~\cite{DeutschPRSA1999}. To attain such a task, projective measurements suffice.

In the case of multiple observables, noncommutativity prevents from obtaining an analogue joint distribution built on projective measurements. A solution was proposed by Wigner, leading to his celebrated function~\cite{Leonhardt}. A so-called quasiprobability distribution is obtained for which negative values can arise, thus representing a signature of genuine quantumness for the state. It has also been linked to the possibility of associating a local hidden-variable model, depending on the specific physical context, with the quantum state~\cite{CETTO1985304,PhysRevLett.82.2009,Delfosse_2017,haferkamp2021,PhysRevLett.129.230401}, as well as to inefficient simulation by classical computers~\cite{MariPRL2012,Veitch_2012,PhysRevX.6.021039,frigerio2025}.

These results have inspired the search for protocols to describe the statistics of outcome pairs returned from measuring noncommuting observables at multiple times~\cite{johansen2007quantum,HofmannPRL2012,HofmannNJP2014,piacentini2016measuring,ArvidssonShukurJPA2021}. In the case of two sequential measurements, often called two-point measurement (TPM) scheme in quantum thermodynamics~\cite{campisi2011colloquium,EspositoRMP2009}, commensurate quasiprobabilities (CQs) have been defined to capture the contextuality of the measurements' action via the presence of negative values~\cite{PhysRevA.88.052123,PhysRevA.96.042121}. Negativity of CQs arises from the state-collapse induced by the first measurement of the sequence, causing a modification of the coherence properties of the initial state that conversely are erased by the TPM scheme.

As proved in \cite{LostaglioQuantum2023}, going beyond the TPM scheme entails that a unique description of the outcome statistics at distinct times is no longer valid, as long as the observables do not commute. In this regard, descriptions using quasiprobabilities, alternative to CQs, have been adopted to fully account for the incompatibility between measurement observables and the initial state~\cite{Solinas2016probing,Hofer2017quasi,lostaglio2018quantum,SolinasPRA2022,FrancicaPRE2022_1,FrancicaPRE2022_2}. A case in point is the Kirkwood-Dirac quasiprobability~\cite{yunger2018quasiprobability,LostaglioQuantum2023,GherardiniTutorial,ArvidssonShukur2024review}, and its real part known as the Margenau-Hill quasiprobability (MHQ)~\cite{margenau1961correlation}.

In all these approaches, measurements are typically taken as projective operations in Von Neumann's terms~\cite{Ryu2019}, but they result in a collapse of the system's wave-function, hence in the disruption of the corresponding quantum state. On the other hand, weak measurements are a viable option to carry out {\it sequential measurements}~\cite{PhysRevLett.60.1351}. These can be interpreted as coarse-grained measurements~\cite{KoflerPRL2007}, in that their outcomes cannot unambiguously distinguish between orthogonal states of the system. Despite this, averaging over many instances still yields the correct expectation value~\cite{PhysRevLett.92.190402,PhysRevLett.94.220405,PhysRevA.65.032111,PhysRevA.73.012113}.
The resulting loss of information can be related to the violation of macrorealism in Legget-Garg inequalities~\cite{PhysRevLett.100.026804,Suzuki_2012,ClementePRA2015}, and, based on anomalous values arising in post-selection, to a witness of contextuality~\cite{BednorzPRL2010,PhysRevLett.113.200401,PhysRevLett.116.180401,Cimini_2020,PhysRevLett.126.100403}. Although the original definitions of quasidistributions~\cite{PhysRevA.88.052123} are suited to generalized measurement, virtually all investigations have only considered projective measurements, making the physical picture of quasiprobabilities based on weak measurements almost unexplored. Further, while the use of generalized measurements has received some attention~\cite{ArvidssonShukur2024review}, weak measurements deserve a special place since they preserve an intuitive physical meaning and are characterized by a parameter tuning over different regimes.

In this article, we describe theoretically weak variants of CQs and MHQs, in discrete quantum systems with arbitrary, finite dimension. Further, we obtain them in an experiment on weak measurements of qubits coded in the polarisation of single photons, retrieved from the direct implementation of the {\it weak-sequential measurement scheme}, shown in Fig.~\ref{fig:setup}a. In such a scheme the first standard strong measurement is replaced by a positive operator-valued measure (POVM), expressed as a function of a parameter tuning from strong to a weak regime. The weak-CQ and weak-MHQ distributions so introduced set a natural way to draw the line between the incompatibility of measurement observables, and measurement coarse-graining. The marginals of weak-CQ and weak-MHQ distributions retain their physical meaning, also encompassing features and limitations of the weak measurement scheme. These concepts are demonstrated in an experiment using the weak measurement of photonic qubits.

\begin{figure}[t!]
    \centering
    \includegraphics[width=\columnwidth]{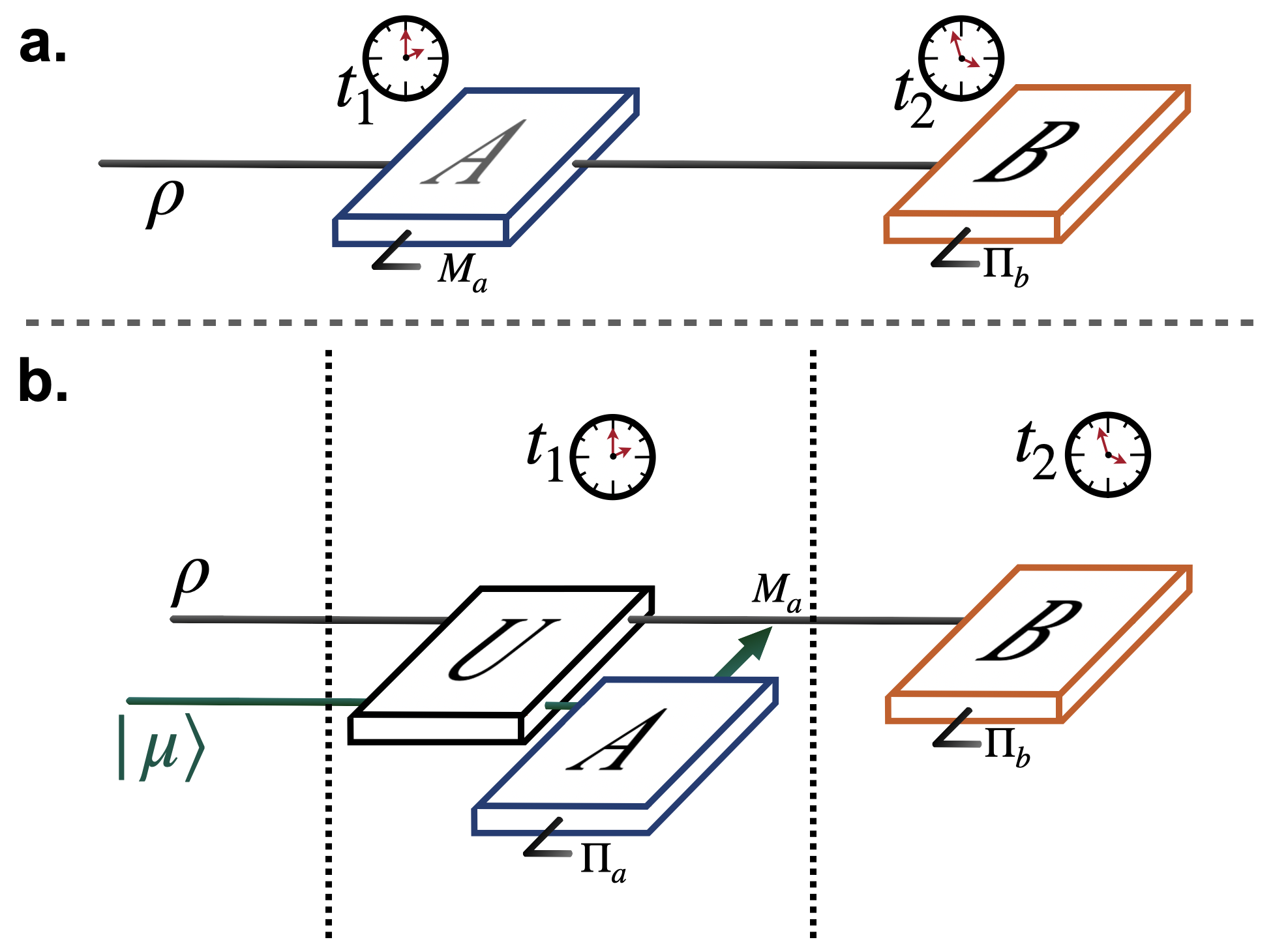}
    \caption{Weak-sequential measurements on a quantum system initialized in a generic density operator $\rho$. a) Pictorial representation of the two-time measurement scheme where the first measurement associated to the observable $A$ is carried out by a POVM with elements $M_a$, and the second is a standard strong measurement of $B$ by means of projectors $\Pi_b$. b) Conceptual approach to the implementation of the weak-sequential measurement scheme. The unitary operator $U$ couples the system, initially in the state $\rho$ (black line), with a pointer in the state $\ket{\mu}$ (green line). Following this operation, a strong projective measurement of $A$ on the pointer can yield information about the system, depending on the initial preparation $\ket{\mu}$. The system thus is still available for further measurements at later times.} 
    \label{fig:concept}
\end{figure}

\section{Background}

We consider a generic density operator $\rho$ describing a quantum state, undergoing the process in Fig.~\ref{fig:concept}a. At the time $t_1$ the value $a$ of the observable $A=\sum_{a}a\,\Pi^{A}_{a}$ is recorded, thus collapsing the wave-function onto the corresponding eigenvector. At a later time, $t_2>t_1$ a measurement of $B=\sum_{b}b\,\Pi^{B}_{b}$ (with $[A,B]\neq 0$) yields the value $b$. It can also encompass in-between quantum dynamics by replacing $\Pi^{B}_{b}$ with the corresponding time-evolved operator in the Heisenberg picture, e.g., $\widehat{\Pi}_b^B = e^{iH(t_2-t_1)/\hbar}\Pi_b^Be^{-iH(t_2-t_1)/\hbar}$ for unitary dynamics governed by a Hamiltonian $H$~\cite{GherardiniTutorial}. As a result, the joint probability of recording these two measurement outcomes is 
\begin{equation}\label{eq:properjoint}
    p(a,b) \equiv {\rm Tr}\left[ \Pi^{B}_{b} \Pi^{A}_{a} \rho \Pi^{A}_{a} \right],
\end{equation}
which is returned by the TPM scheme~\cite{campisi2011colloquium}.

The joint probabilities $p(a,b)$ form a valid distribution that is nevertheless insensitive to initial coherences in the eigenstates of $A$ due to the invasivity of the first measurement. Consequently, the marginal $\sum_{a}p(a,b)$ is generally different from $p_{\rm fin}(b)={\rm Tr}[\Pi^{B}_{b}\rho]$, i.e., the probability of obtaining the value $b$ for the observable $B$ directly on the initial state $\rho$. The definition of the CQ, originally introduced in~\cite{PhysRevA.88.052123,PhysRevA.96.042121,Ryu2019}, makes use of the unmeasured state’s statistics, in order to account for the disturbance caused by the measurement of $A$:
\begin{equation}\label{eq:Wstrong}
q_{\rm C}(a,b) \equiv p(a,b) +
\frac{1}{d}\Big( p_{\rm fin}(b) - p_{\rm post}(b) \Big),
\end{equation}
where $p_{\rm post}(b) \equiv \sum_{a}p(a,b) = \sum_{a}{\rm Tr}\left[ \Pi^{B}_{b} \Pi^{A}_{a} \rho \Pi^{A}_{a} \right] = \sum_{a}p_{\rm in}(a){\rm Tr}\left[ \Pi^{B}_{b} \Pi^{A}_{a} \right]$ is the marginal distribution of the TPM joint probability over $a$, with $p_{\rm in}(a) = {\rm Tr}[\Pi^{A}_{a} \rho]$ and $d$ denoting the dimension of the system. In Eq.~\eqref{eq:Wstrong}, the term in the parenthesis quantifies how much $p_{\rm post}(b)$ differs from $p_{\rm fin}(b)$, computed on the initial density operator $\rho$.

Refs~\cite{PhysRevA.88.052123,PhysRevA.96.042121} demonstrated CQs are valid quasidistributions. These are linear in $\rho$, sum to 1 and their marginal distributions return the unperturbed probabilities $p_{\rm fin}(b)$, $p_{\rm in}(a)$  of measuring $B$ and $A$ separately. While they can be rigorously grounded in a suitable characteristic function for the sequential measurement scheme, the definition \eqref{eq:Wstrong} is operational, and easier to interpret~\cite{Ryu2019}. When $[\rho,A]=0$, the joint probability of the TPM scheme is recovered. If $[\rho,A]\neq0$, instead, the quantities $q_C(a,b)$ may attain negative values. It has been established this is a manifestation of the failure of non-contextual models at describing the experiment~\cite{PhysRevA.88.052123}.

While the CQs amend the TPM probability distribution by including measurement disturbance, MHQ directly identifies observables incompatibility~\cite{LostaglioQuantum2023,GherardiniTutorial,ArvidssonShukur2024review}. The two concepts are intimately linked to each other. For rank-1 projectors $\Pi^{A}_{a} = \ketbra{a}{a}$, $\Pi^{B}_{b} = \ketbra{b}{b}$, MHQ is defined as
\begin{equation}\label{eq:MHQ}
   q_{\rm MH}(a,b) \equiv \text{Re}\left[ {\rm Tr}\left[ \Pi^{B}_{b} \Pi^{A}_{a} \rho \right] \right] = \left\vert\braket{b}{a}\right\vert^2 \text{Re}\left[ \frac{\bra{b}\rho\ket{a}}{\braket{b}{a}} \right].
\end{equation}
By construction, $q_{\rm MH}(a,b)$ are real numbers that sum to 1, returning the correct marginal distributions for both $A$ and $B$, and are linear in $\rho$. Moreover, being quasiprobabilities, $q_{\rm MH}(a,b)$ can also assume negative values. The latter scenario explicitly targets the incompatibility of $\rho$ and $A$, given that MHQs are equal to the joint probabilities in Eq.~\eqref{eq:properjoint} of the TPM scheme for $[\rho,A] = 0$.

\section{Weak-quasiprobabilities}

Replacing the strong measurement of $A$ at time $t_1$ by its weak version, implemented by a general POVM with elements $M^A_a$, amounts to introduce coarse-graining~\cite{KoflerPRL2007}. This may lead to a classical description of the outcomes’ statistics in that they can be modelled by a positive two-time probability distribution.

Let us refer to the scheme detailed in Fig.~\ref{fig:concept}. At $t_1$ the system in the state $\rho$ is coupled to an ancillary pointer state prepared in the state $\ket{\mu}$ with the same dimensionality $d$ as the system. The coupling between the system and the pointer is realized by means of a modular unitary $U$ defined on the system computational basis $\ket{a}$ as 
\begin{equation}
\label{eq:unitary}
U = \sum_{a=0}^{d-1} \ketbra{a}{a} \otimes V^a, 
\end{equation}
where $V$ is such that $V\ket{a}=\ket{a\oplus 1}=\ket{a+1{\mod d}}$, $V^0=\mathbb{I}$, and $V^n = V\cdot V^{n-1}$ for $n>0$.

We can illustrate how the scheme works by first considering the case of a pure input: $\rho=|\psi\rangle\langle
\psi|$ with $\ket{\psi}=\sum_a c_a\ket{a}$. When $\ket{\mu} = \ket{0}$, then 
\begin{equation}
U\ket{\psi}\!\ket{0} = \sum_a c_a \ket{a}\!\ket{a}.
\end{equation} 
In these conditions, a strong measurement of the observable $A$ on the pointer would yield the outcome $a$ with probability $|c_a|^2$. Therefore, this operation realizes a measurement of the observable $A$ on the system. Instead, $\ket{\mu_0} \equiv \frac{1}{\sqrt{d}} \sum_a \ket{a}$ has the property that $V\ket{\mu_0}=\ket{\mu_0}$; hence, 
\begin{equation}
U\ket{\psi}\!\ket{\mu_0} = \ket{\psi}\ket{\mu_0}.
\end{equation}
Any projective measurement on the pointer bears no implication for the system. By interpolating between the two cases with $|\mu\rangle=|0\rangle$ or $|\mu_0\rangle$, we can realize a weak measurement of $A$ with arbitrary strength. Indeed, taking the initial state of the pointer as 
\begin{equation}
\ket{\mu} = u_0\ket{0} + \frac{u_1}{\sqrt{d-1}}\sum_{a\neq 0}\ket{a}
\end{equation}
with real coefficients $u_0$ and $u_1$, one obtains 
\begin{equation}
U\ket{\psi}\!\ket{\mu} = \omega_0 \sum_a c_a\ket{a}\!\ket{a} + \omega_1\ket{\psi}\!\ket{\mu_0},
\label{eq:endmatter}
\end{equation}
where $\omega_0 \equiv u_0 - u_1/\sqrt{d-1}$ and $\omega_1 \equiv u_1 \sqrt{d/(d-1)}$. Thanks to the correlations introduced by the unitary $U$, from Eq.~\eqref{eq:endmatter} we have that a strong measurement of the observable $A$ on the pointer can realize a weak measurement of the same observable $A$ on the system. At $t_2$ the observable $B$ is measured with a standard strong measurement. Therefore, the probabilities of the weak-sequential measurement scheme in Fig.~\ref{fig:setup} are given by
\begin{eqnarray}\label{eq:p_weak}
    &&p_{\rm weak}(a,b) \equiv \left\vert \bra{b}\!\bra{a}U\ket{\psi}\!\ket{\mu} \right\vert^2 = \nonumber \\
    &&\omega_0^2 \vert c_a \vert^2 \vert \braket{b}{a} \vert^2+\frac{\omega_1^2}{d}\vert\braket{b}{\psi} \vert^2+\frac{2\omega_0\omega_1}{\sqrt{d}}
    \text{Re}\left[c_a \braket{b}{a}\braket{\psi}{b}\right] \nonumber \\
\end{eqnarray}
where the normalization $\omega_0^2+\omega_1^2+2\omega_0\omega_1/\sqrt{d}=1$ holds.

By linearity, this formulation generalizes for a
generic mixed state input $\rho$, whereby the state of the system after the measurement is $\left(m_a^A\right)^\dag\rho \, m_a^A/{\rm Tr}\big[(m_a^A)^\dag m_a^A\big]$ with $m_a^A = \omega_0\Pi_a^A + \omega_1 \mathbb{I}/\sqrt{d}$. The latter realizes a $d$-outcome POVM with elements~\cite{PhysRevLett.104.080503,Mancino18} 
\begin{equation}
M_a^A=\left(m_a^A\right)^\dag m_a^A = (1-\omega_1^2)\Pi_a^A+\omega_1^2 \frac{\mathbb{I}}{d}.
\end{equation}
The measurement strength is captured by the parameter $K=1-\omega_0^2$, ranging from $K=0$ for no measurement to $K=1$ for a fully projective measurement. The quantity K can be identified as a formalisation of which-path information in interferometry~\cite{PhysRevLett.77.2154,PhysRevA.73.012113}. In this general setting, the joint probability of the weak-sequential measurement scheme satisfy the relation
\begin{equation}\label{eq:p_weak_rho}
p_{\rm weak}(a,b) = \omega_0^2 p(a,b) + \frac{\omega_1^2}{d} p_{\rm fin}(b) + \frac{2\omega_0\omega_1}{\sqrt{d}}q_{\rm MH}(a,b).
\end{equation}


We can now elaborate on the definition of a new class of quasiprobabilities suited to such weak sequential schemes. A weak version of the CQs demand substituting $p(a,b)$ with $p_{\rm weak}(a,b)$ in the definition of Eq.~\eqref{eq:Wstrong}: 
\begin{equation}\label{eq:WeakCQ_definition}
    \tilde q_{\rm C}(a,b) = p_{\rm weak}(a,b) + \frac{1}{d}\Big( p_{\rm fin}(b) - \sum_{a}p_{\rm weak}(a,b) \Big).
\end{equation}
The expression in \eqref{eq:WeakCQ_definition} can be related more clearly to the ordinary quasiprobabilities by using Eq.~\eqref{eq:p_weak_rho} and the fact that $p_{\rm fin}(b)-\sum_a p_{\rm weak}(a,b) = \omega_0^2\left(p_{\rm fin}(b)-p_{\rm post}(b)\right)$:
\begin{equation}\label{eq:weakCQ}
    \tilde q_{\rm C}(a,b) = \omega_0^2 q_C(a,b) + \frac{ \omega_1^2 }{d} p_{\rm fin}(b) + \frac{2\omega_0\omega_1 }{\sqrt{d}}q_{\rm MH}(a,b).
\end{equation}
Importantly, $\tilde q_{\rm C}(a,b)$ sum to $1$, have $p_{\rm fin}(b)$ as the marginal when summing over $a$, and are so that
\begin{equation}\label{eq:marginal_b_weak_CQ}
    \sum_{b}\tilde q_{\rm C}(a,b) = \sum_{b}p_{\rm weak}(a,b) = K p_{\rm in}(a) + \frac{(1-K)}{d}
\end{equation}
from marginalizing over $b$. This represents the expected marginal allowed by the weak measurement, thus supporting the validity of $\tilde q_{\rm C}(a,b)$ as meaningful quasiprobabilities.

As for weak-MHQs, we replace in Eq.~\eqref{eq:MHQ} the projector $\Pi_a^A$ with its weak version $M_a^A$ that yields to
\begin{equation}\label{eq:weak-MHQ}
\tilde q_{\rm MH}(a,b) = K \, q_{\rm MH}(a,b) + \frac{ (1-K) }{d}p_{\rm fin}(b)\,.
\end{equation}
Marginalizing $\tilde q_{\rm MH}(a,b)$ over the outcomes $a$ or $b$ returns the same marginal distributions as for the weak-CQs $\tilde q_{\rm C}(a,b)$. For qubit systems ($d=2$), $\tilde q_{\rm MH}(a,b) = \tilde q_{\rm C}(a,b)$ for all strengths.

These weak quasidistributions can attain negative values, and their meaning remains the same for the their counterparts based on projective measurements: they now also weight-in the effect of coarse-graining on the possibility of using a classical or quantum probability description. Considering the weak-MHQ in \eqref{eq:weak-MHQ}, it is easy to identify a limit value $\overline{K}(a,b)$ below which the single instance $q_{\rm MH}(a,b)$ is non negative:
\begin{equation}\label{eq:K_threshold}
    \overline{K}(a,b) = \frac{1}{1-\frac{d\, q_{\rm MH}(a,b)}{p_{\rm fin}(b)}}\,.
\end{equation}
Any coarse-graining of the measurements of $A$ such that $K < \overline{K}$ makes the experiment compatible with a classical, noncontextual description.

\section{Experiment and Results}

\begin{figure}[t!]
    \centering
    \includegraphics[width=\columnwidth]{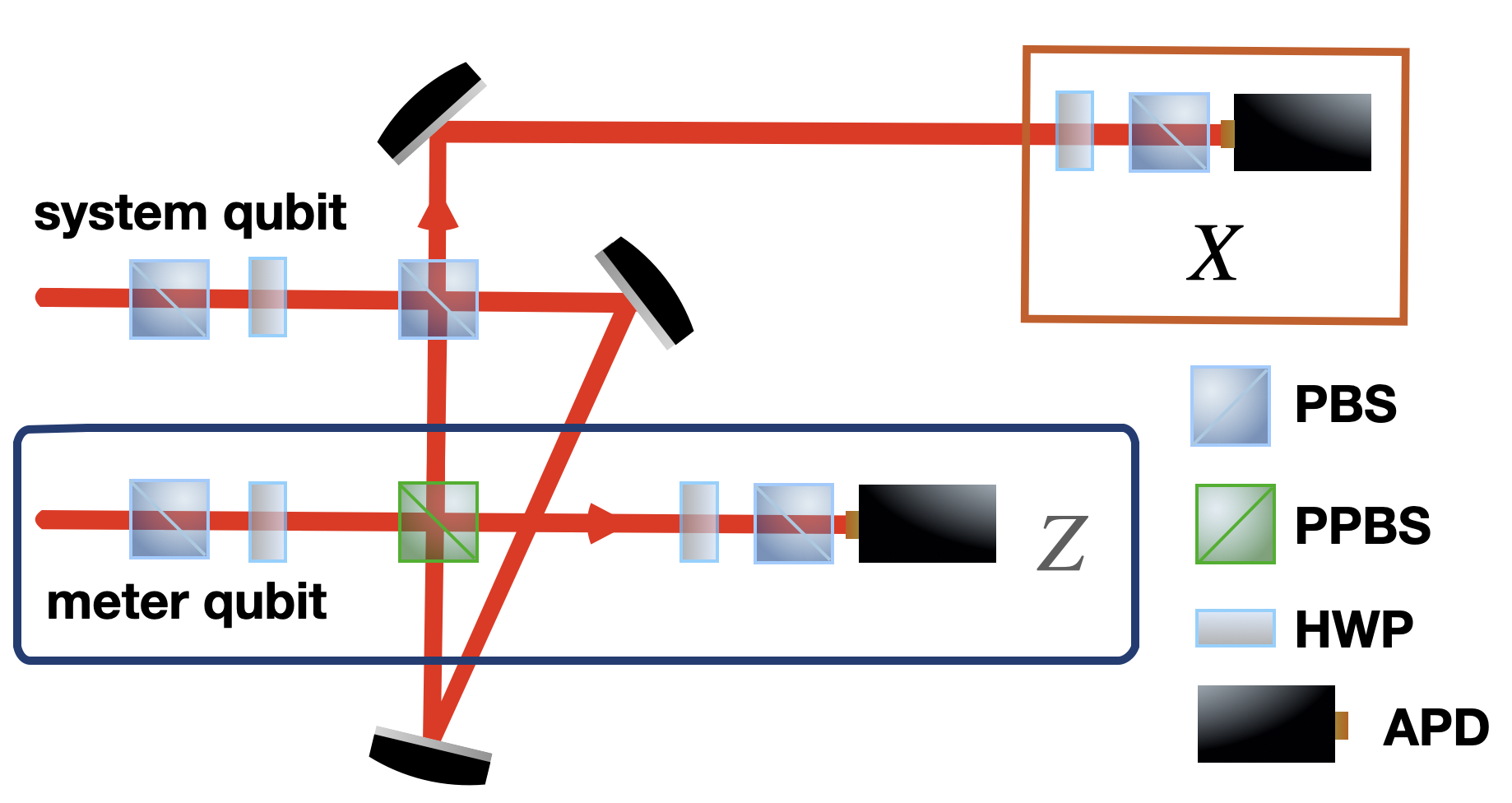}
    \caption{Setup for the experimental test with photon qubits. Photon pairs are produced by a parametric down-conversion source (3 mm barium borate crystal pumped by a 100 mW cw laser at 405 nm), and delivered to a control-Z gate by single-mode fibres. The gate uses a partially polarising beam splitter (PPBS) with transmissions $T_H=1$, $T_V=1/3$ embedded within a Sagnac interferometer~\cite{Cimini_Thermo}. Polarisation-sensitive interference and post-selection of coincidence events result in the control-Z operation. Linear polarisation states are prepared and measured by means of half-waveplates (HWPs) and polarisation beam splitters (PBSs). Photons are finally sent on fibre-coupled avalanche photo-diodes (APDs).} 
    \label{fig:setup}
\end{figure}

\begin{figure}
    \centering
    \includegraphics[width=\columnwidth]{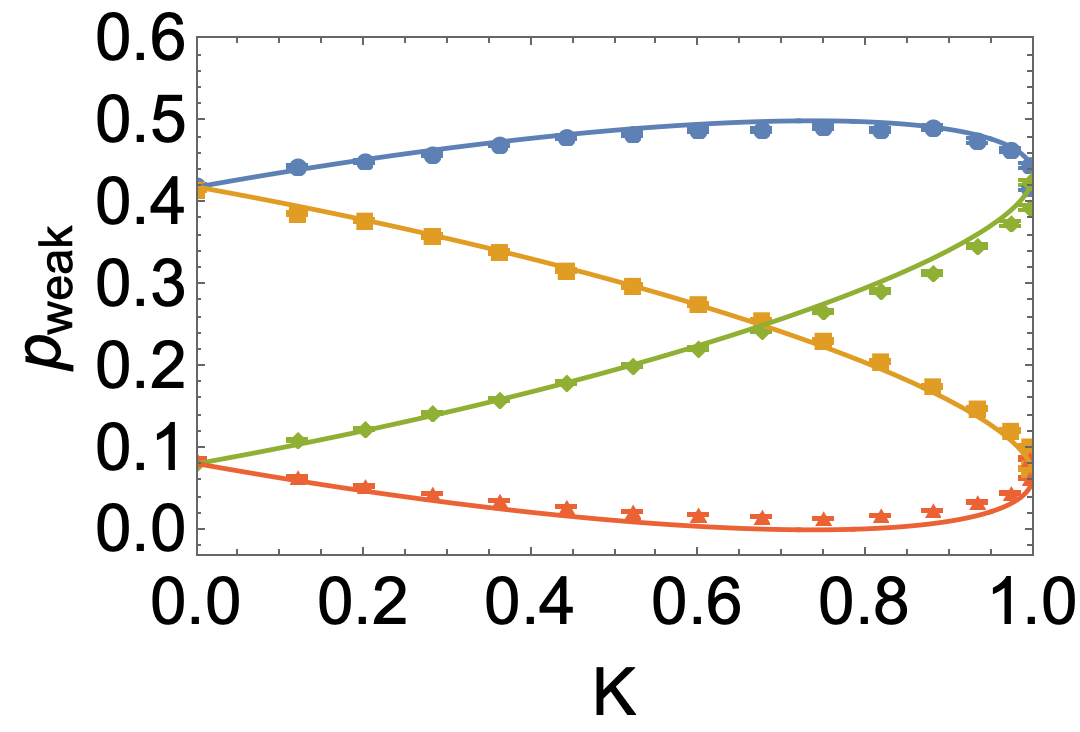}
    \caption{Experimental probabilities $p_{\rm weak}(a,b)$, from the weak measurement device depicted in Fig.~\ref{fig:setup}, as a function of the measurement strength $K$. In the figure, $(a,b)=(H,D)$ (blue), $(V,D)$ (yellow), $(H,D^\perp)$ (green), $(V,D^\perp)$ (red) for the input wave-function $\ket{\psi}=\cos(2\theta_0)\ket{H}+\sin(2\theta_0)\ket{V}$ ($\theta_0=10.6^\circ$). The solid curves denote the theoretical prediction, and the points are the experimental data.}
    \label{fig:prob}
\end{figure}

We realized the weak-sequential scheme, depicted in Fig.~\ref{fig:setup} on photonic qubits coded in the polarisation of single photons from parametric down conversion. The system is prepared in the superposition $\cos(2\theta_0)\ket{H}+\sin(2\theta_0)\ket{V}$, where we use the standard notation for the horizontal and vertical polarisations rotated by the angle $2\theta_0$. The first measurement of the Pauli observable $A=Z$ is weak and is realized by means of a control-Z gate, with a second qubit acting as its target~\cite{PhysRevLett.94.220405}; our implementation corresponds to taking $V=Z$ in the coupling unitary \eqref{eq:unitary}. The initial diagonal pointer state $\ket{0}=\ket{D}=\left(\ket{H}+\ket{V}\right)/\sqrt{2}$ leads to a full-strength measurement ($K=1$), while $\ket{\mu_0}=\ket{H}$ corresponds to no measurement. The logical states of the two qubits, system and pointer, are encoded as different polarisation bases. Hence, the most general state we employ to tune the measurement strength $K$ is 
\begin{equation}
    \ket{\mu}=\cos(2\phi)\ket{D}+\sin(2\phi)\ket{D^\perp},
\end{equation}
with $\ket{D^\perp}=\left(\ket{H}-\ket{V}\right)/\sqrt{2}$  and $K=2\cos^2(2\phi)-1$~\cite{PhysRevLett.77.2154}. The pointer is projected into the basis defined by $\{ \ket{D}, \ket{D^\perp} \}$, while either the POVM element $M_0^Z = K\ketbra{H}{H}+(1-K)\,\mathbb{I}/2$ or $M_1^Z = K\ketbra{V}{V}+(1-K) \, \mathbb{I}/2$ is applied to the system. The latter is projected by a conventional polarisation measurement to assess the Pauli observable $B=X$. The corresponding experimental probabilities of the weak-sequential measurement scheme are reported in Fig.~\ref{fig:prob}, and have served as the starting point for our analysis. The values of $p_{\rm fin}(b)$ are obtained by running the measurement at $K = 0$, and then tracing out the pointer.

\begin{figure}
    \centering
    \includegraphics[width=\columnwidth]{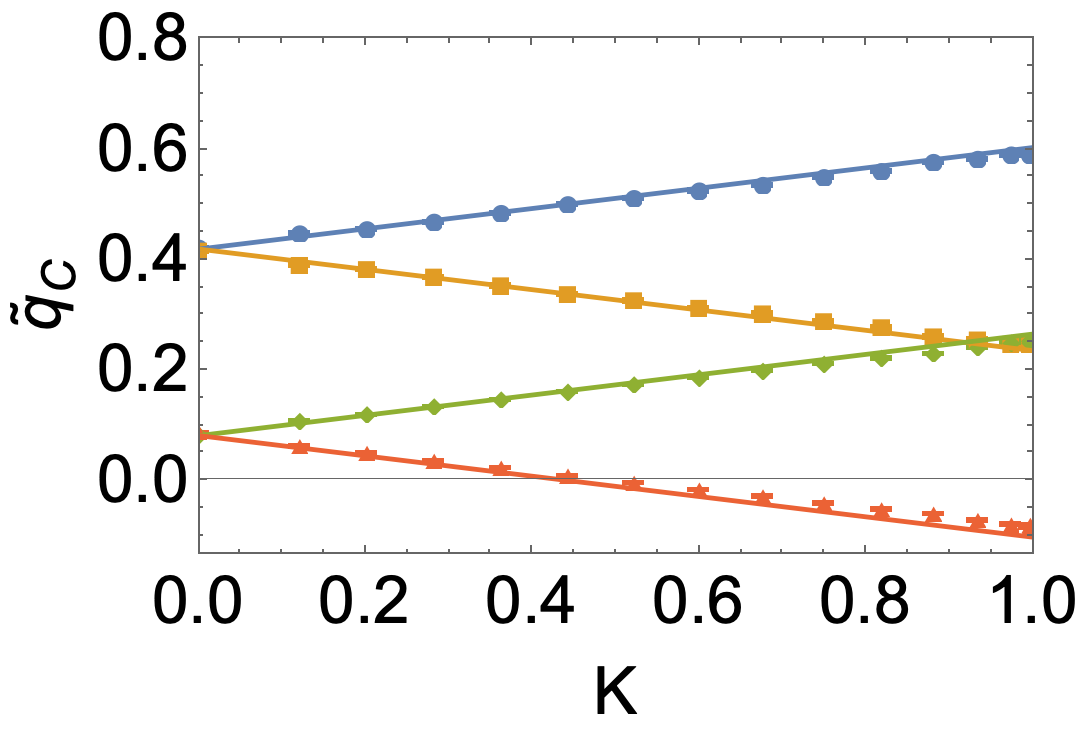}
    \caption{Weak-quasiprobabilities $\tilde q_{\rm C}(a,b)$, with $(a,b)=(H,D)$ (blue), $(V,D)$ (yellow), $(H,D^\perp)$ (green), $(V,D^\perp)$ (red) for the input wave-function $\ket{\psi}=\cos(2\theta_0)\ket{H}+\sin(2\theta_0)\ket{V}$ ($\theta_0=10.6^\circ$), as a function of the measurement strength $K$. The solid curves denote the theoretical predictions, while the points are the experimental data. The size of the error bars is of the order of $2\cdot 10^{-3}$, hence not visible on the plot.}
    \label{fig:CQ}
\end{figure}

In Fig.~\ref{fig:CQ} we plot the four possible weak-CQs $\tilde q_{\rm C}(a,b)$ as a function of the measurement strength $K$, as obtained by the definition in Eq.~\eqref{eq:WeakCQ_definition} using the probabilities $p_{\rm weak}(a,b)$ retrieved from the experiment. The data refer to $\theta_0=10.6^\circ$ that corresponds to the almost maximal negativity in the ordinary CQs $q_{\rm C}(a,b)$. The weak quasiprobability can show a negative part, and its origin can be traced back to two effects: the incompatibility of the observables $A$ and $B$ (in general, non-commuting operators), and the presence of quantum coherence in the initial state. For qubits both effects are quantified by the MHQs $q_{\rm MH}(a,b)$. Conversely, for a quantum system with generic dimension, the incompatibility of the observables and quantum coherence in the initial state can manifest in a distinct way, and are associated with the (strong) CQ and MHQ distributions, respectively. From the figure, we can thus observe the effect of the coarse-graining on the (first) measurement of $A$, which we experimentally obtain by decreasing $K$. There is indeed a threshold value of $K$ under which all $\tilde q_{\rm C}(a,b)$ become positive, and a classical description of the data becomes attainable.

\begin{figure}
    \centering
    \includegraphics[width=\columnwidth]{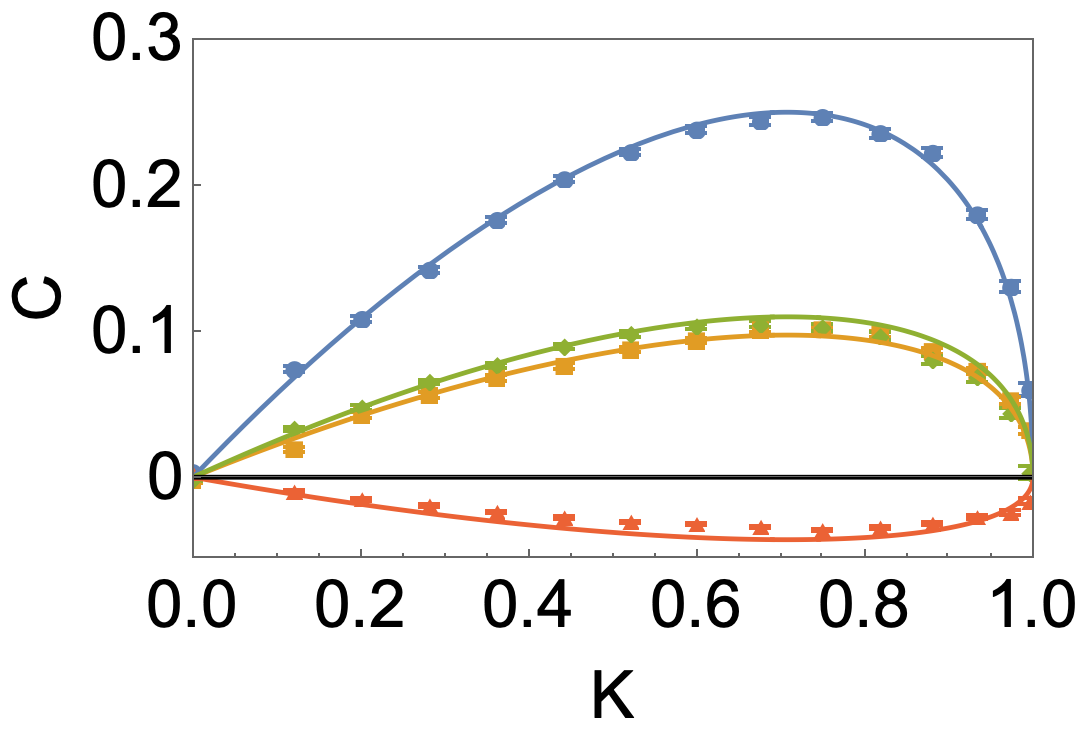}
    \caption{Coherent content $C(a,b)$ detected by $p_{\rm weak}(a,b)$, with $(a,b)=(H,D)$ (blue), $(V,D)$ (yellow), $(H,D^\perp)$ (green), $(V,D^\perp)$ (red) for the input wave-function $\ket{\psi}=\cos(2\theta_0)\ket{H}+\sin(2\theta_0)\ket{V}$ ($\theta_0=10.6^\circ$), as a function of the measurement strength $K$. The solid curves denote the theoretical predictions, while the points are the experimental data. Discrepancies can be attributed mostly to the imperfect non-classical interference contrast at the gate. Errors on data are calculated by means of a Monte Carlo procedure, assuming the measured counts follow the Poisson distribution. These errors are quite small and not visible on the plot.}
    \label{fig:Coherence}
\end{figure}

The expression for the weak quasiprobability \eqref{eq:p_weak_rho} spells out that each $p_{\rm weak}(a,b)$ is constituted by three contributions with an intuitive physical meaning. In particular, the two terms $p(a,b)$ and $p_{\rm fin}(b)/d$ capture the extreme cases, in which a full-strength measurement of $A$ is carried out and no measurement of $A$ is performed, respectively. Under those conditions, the coherence of $\rho$ in the eigenbasis of $A$ is irrelevant. The third term $C(a,b) = (2\omega_0\omega_1/\sqrt{d}) \, q_{\rm MH}(a,b)$, thus, identifies the impact of such coherence in the joint statistics of $A$'s and $B$'s outcomes, given that part of it is preserved by the weak measurement. Using Eq.~\eqref{eq:p_weak_rho}, we measure it as $C(a,b) \equiv p_{\rm weak}(a,b) - \omega_0^2 p(a,b) -(\omega_1^2 / d) p_{\rm fin}(b)$, where the three quantities are obtained in separated experiments ran at strengths $K$ equal to $1$ and $0$, respectively. Fig.~\ref{fig:Coherence} reports the measured values as a function of the measurement strength $K$, together with the theoretical predictions. We observe that in any experimental routine we performed, quantum coherence in the initial state affects the statistics originating from $A$ and $B$, in terms of MHQs. In the experiments, we are able to detect it directly from implementing the weak-sequential measurement scheme of Fig.~\ref{fig:setup}. In Fig.~\ref{fig:Coherence}, it is also shown that $C(a,b)$ can take both positive and negative values, as the real part of a quantum coherence does. Interestingly, being $C(a,b)$ directly proportional to $q_{\rm MH}(a,b)$, the negative parts of $C(a,b)$ witness the negativity of CQ and MHQ distributions, as well as of their weak counterparts (see also Fig.~\ref{fig:CQ}). 

The weak-sequential measurement scheme and in particular Eq.~\eqref{eq:p_weak_rho} provide an alternative way to reconstruct MHQs, with respect to the weak-TPM scheme~\cite{johansen2007quantum,LostaglioQuantum2023,hernandez2022experimental,GherardiniTutorial}, and brings suggestions on how to extend to weak measurements the use of quasiprobabilities.
The weak-TPM scheme relies on the relation
\begin{equation}\label{eq:weak-TPM_relation}
    q_{\rm MH}(a,b) = p(a,b) + \frac{1}{2}\Big( p_{\rm fin}(b) - w(a,b) \Big).
\end{equation}
In Eq.~\eqref{eq:weak-TPM_relation}, $w(a,b) \equiv {\rm Tr}\left[ \Pi^{B}_{b}\rho_{\rm NS}(a) \right]$ denotes the weak-TPM joint probabilities, where $\rho_{\rm NS}(a) \equiv \Pi^{A}_{a} \rho \Pi^{A}_{a} + ( \mathbb{I} - \Pi^{A}_{a} )\rho( \mathbb{I} - \Pi^{A}_{a} )$ is the density operator returned by performing non-selective projective measurements with projectors $\Pi_a$ and $\Pi^{\perp}_a \equiv \mathbb{I} - \Pi^{A}_{a}$. Interestingly, for qubits ($d=2$), Eq.~\eqref{eq:weak-TPM_relation} coincides with Eq.~\eqref{eq:Wstrong}.\\ 
The advantage of using Eq.~\eqref{eq:p_weak_rho} instead of realizing the weak-TPM scheme lies in the fact that it is no longer necessary to implement projectors such as $\Pi^{\perp}_a$ that are more demanding in terms of the preparation and the manipulation of ancillary pointers. The normalization condition ${\omega_0^2+\omega_1^2+2\omega_0\omega_1/\sqrt{d}=1}$ guarantees that the MHQ distribution got from Eq.~\eqref{eq:p_weak_rho} has correct marginals. The weak-sequential measurement scheme resembles the one allowing for the direct reconstruction of the quantum state~\cite{Lundeen2011,PhysRevLett.121.230501,PhysRevLett.127.180401}. Here, the coarse-graining is instrumental to the reconstruction, as it enables access to quantum coherence in the initial state~\cite{PhysRevA.107.022408,budiyono2024quantum}. 

\begin{figure}
    \centering
    \includegraphics[width = \columnwidth]{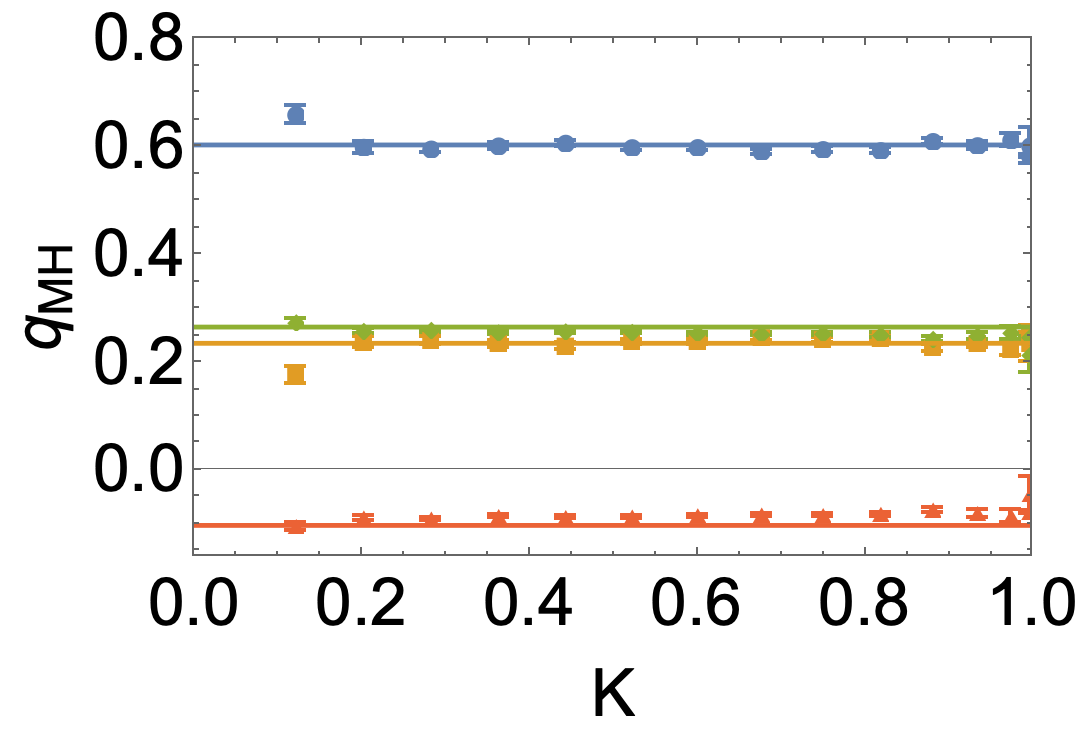}
    \caption{Margenau-Hill quasiprobabilities $q_{\rm MH}(a,b)$, with $(a,b)=(H,D)$ (blue), $(V,D)$ (yellow), $(H,D^\perp)$ (green), $(V,D^\perp)$ (red) for the input wave-function $\ket{\psi}=\cos(2\theta_0)\ket{H}+\sin(2\theta_0)\ket{V}$ ($\theta_0=10.6^\circ$), as a function of the measurement strength $K$. The solid curves denote the theoretical prediction, and the points are the experimental data. Errors are calculated by means of a Monte Carlo procedure, assuming Poisson distribution of the measured counts.}
    \label{fig:MH}
\end{figure}

In Fig.~\ref{fig:MH} we report the experimental reconstruction of $q_{\rm MH}(a,b)$ using Eq.~\eqref{eq:p_weak_rho}, where we have taken the same input quantum state as in Fig.~\ref{fig:CQ}. It is worth noting that more reliable performance are obtained away from the extreme conditions $K=0,1$ in terms of both precision and accuracy. When $K$ is nearly zero, little is learned about $A$, while for $K$ close to $1$ there is small coherence between the eigenstates $\ket{a}$ of $A$ left for the  measurement of $B$. The experimental imperfections then lead to a relatively more severe impact, especially with a small informational content.  

\section{Conclusion}

We have shown how two classes of quasiprobabilities (CQs and MHQs) can be defined when the first measurement in a sequence of measurements is weak. These weak-quasiprobabilities can still attain negative values, linked to contextuality. Realizing a sequential scheme measuring negativity---here the weak-sequential measurement scheme (Fig.~\ref{fig:setup})---can be implemented but at the price of refraining full information on the initial conditions. 

The weak-sequential measurement scheme is an alternative tool for the analysis of extended protocols with sequential measurements at two or multiple times, with a rate of invasivity that one could control by the disappearance of negative values of a weak-quasiprobability. These can be thought of as a complementary description of quantum states at multiple times~\cite{horsman2017,fullwood2024}, and provide a tool for the monitoring of quantum objects with applications to quantum sensing~\cite{PhysRevLett.96.020408,PhysRevLett.104.080503,Blok14,PhysRevLett.114.210801,doi:10.1126/sciadv.adi5261}, and quantum thermodynamic processes~\cite{levy2020quasiprobability,maffei2022anomalous,hernandez2022experimental,Cerisola2023aWigner,SagarQSL2024,GherardiniTutorial,hernandezArXiv2024Interfero}. A concrete example could be the realization of a quantum engine~\cite{CangemiArXiv2023Review} fueled by quantum coherence through a thermodynamic cycle (e.g., in Otto configuration) run on weak measurements.

\section*{Acknowledgements}

We thank R.~Grasland for support during data collection and J. Sperling for valuable discussion. This work was supported by the PRIN project PRIN22-RISQUE-2022T25TR3 of the Italian Ministry of University.
S.G.~acknowledges support from the PRIN project 2022FEXLYB Quantum Reservoir Computing (QuReCo), and the PNRR MUR project PE0000023-NQSTI funded by the European Union--Next Generation EU.
G.B.~is supported by Rome Technopole Innovation Ecosystem (PNRR grant M4-C2-Inv). IG acknowledges the support from MUR Dipartimento di Eccellenza 2023-2027.

\bibliography{main.bib}

\end{document}